\documentclass[conference]{IEEEtran}
\ifCLASSINFOpdf
\else
\fi
\usepackage{mathrsfs}
\usepackage{algorithm} 
\usepackage{algorithmic} 
\usepackage{multirow} 
\usepackage{amsmath}
\usepackage{xcolor}
\usepackage{amssymb}
\usepackage{graphicx}
\usepackage{subfigure}
\usepackage{booktabs}
\usepackage{bm}
\usepackage{cite}
\usepackage{epstopdf}
\usepackage{changepage}
\usepackage{lipsum}
\usepackage{subfigure}
\usepackage{amssymb}
\usepackage{empheq}
\usepackage{amsfonts}
\usepackage{diagbox}
\usepackage{array}
\usepackage{empheq}

\begin{document}
%
\title{Cascade-Net: a New Deep Learning Architecture for OFDM Detection}

\author{Qisheng Huang$^{1}$,
        Chunming Zhao$^{1}$,
         Ming Jiang$^{1}$,
         Xiaoming Li$^{1}$,
         Jing Liang$^{2}$\\
        $^{1}$National Mobile Communications Research Lab., Southeast University, Nanjing 210096, China\\
        $^{2}$Huawei Technologies CO., LTD.\\
        Email:$^{1}$\{qshuang, cmzhao, jiang\_ming, xmli\}@seu.edu.cn, $^{2}$jingliang@huawei.com
        \thanks{}}


%


\maketitle

\begin{abstract}
In this paper, we consider using deep neural network for OFDM symbol detection and demonstrate its performance advantages in combating large Doppler Shift. In particular, a new architecture named Cascade-Net is proposed for detection, where deep neural network is cascading with a zero-forcing preprocessor to prevent the network stucking in a saddle point or a local minimum point. In addition, we propose a sliding detection approach in order to detect OFDM symbols with large number of subcarriers. We evaluate this new architecture, as well as the sliding algorithm, using the Rayleigh channel with large Doppler spread, which could degrade detection performance in an OFDM system and is especially severe for high frequency band and mmWave communications. The numerical results of OFDM detection in SISO scenario show that cascade-net can achieve better performance than zero-forcing method while providing robustness against ill conditioned channels. We also show the better performance of the sliding cascade network (SCN) compared to sliding zero-forcing detector through numerical simulation.

\end{abstract}



%
\IEEEpeerreviewmaketitle

\section{Introduction}
Orthogonal frequency division multiplexing (OFDM) has been widely applied in modern communication. By using fast fourier transformation (FFT), this technique can support high data-rate transmission and achieve high spectral efficiency in wireless communications. Recently, application of this technique in the 5th generation (5G) wireless communication system has been confirmed\cite{ofdm5G}. However, this attractive technique is very sensitive to the carrier frequency offset, phase noise, timing offset, and Doppler spread\cite{iterative}, which can break the orthogonality between subcarriers and cause inter-carrier interference (ICI). With growth of the carrier frequency used in future broadband wireless access and the speed of modern vehicles, the Doppler spread of the wireless channel strongly increases which leads to more severe ICI. To deal with this emerging equalization problem, we propose a deep learning based method which usually performs better than the classical methods, such as zero-forcing and MMSE detection.

During the past few years, with the development of deep learning approach, and the deep learning algorithm, neutral network architecture has been successfully used in the field of computer vision and language processing, given its expressive capacity and convenient optimization capability\cite{Wang2017Deep}. Particularly, many practical deep learning models for physical layer communication come out, such like deep channel estimation\cite{Ye2018Power}, deep channel decoding\cite{deepdecode1}\cite{deepdecode2}, deep MIMO detection\cite{Samuel2018Learning} and autoencoder for deep modulation\cite{Kim2018Deep}\cite{Li2018Achievable}. These specific applications of deep learning in communications can be roughly divided into two types. First, using the deep unfolding\cite{deepunfolding} to add trainable parameters to the classical methods, through this data driven detection to find a promoted algorithm. Second, substituting the modified dense neutral network, convolutional neutral network or residual neutral network for the appropriate parts in communications for enhancement. In this paper, we mainly focus on the first usage and enhance the performance of OFDM systems by improving the classical detection. As the primary requirement of higher transmitting rate, the neutral networks are usually trained off-line and directly applied online with reconfigurable hardware.

The main contribution of this paper is to propose a new architecture in OFDM detection. As mentioned before, Doppler spread can cause severe ICI between subcarriers especially when information are transmitted on high frequency. Inspired by deep MIMO detection\cite{Samuel2018Learning}, we use similar deep unfolding method to create a trainable network through modifying ML algorithm in purpose of combating against ICI. However, training of deep neutral network can easily get into a saddle point or local minimum\cite{deeplearning}, which leads to poor performance in high signal-to-noise rate. In this paper, we creatively propose a cascade structure to handle this issue. In cascade network, neural network is cascaded to a zero-forcing preprocessor being trained and used as a whole part. The thought of cascade network is similar to transfer learning. By adding parameter-fixed network to a new net, the difficulty of training a new high-dimension network to converge sharply decreases. In multi subcarrier scenario, we propose a sliding structure\cite{fenkuai}\cite{Farsad2018Neural}. Our sliding structure consists of two parts: output area (OA) and guarding area (GA). Through careful analysis of adjacent subcarriers' ICI to the subcarrier being detected, we give out the empirical formula used for designing the length of GA and OA. This sliding structure ensures the detecting performance of our cascade-net without adding too much calculation complexity.

\section{SYSTEM MODEL AND NEURAL NETWORK FOR DETECTION }
\subsection{SYSTEM MODEL }
 In our paper we mainly forcus on OFDM detection in single in single out (SISO) scenario. For convenience, we use frequency-domain model to describe the whole system. Assuming that cyclic prefix (CP) is long enough to eliminate internal symbol interference, transmitting and detection of each OFDM symbol would be independent. Considering an OFDM system with $N$ subcarriers, the transmitted data in time-varying multipath channel is  $x(n)$. When the transmitted signal passes through the channel $h(n,l)$, the received signal can be represented as\cite{novel}
\begin{equation}
\begin{aligned}
  y\left( n \right)&=h\left( n,l \right)*x\left( n \right)+w\left( n \right) \\
\end{aligned}
\end{equation}
where $*$ denotes the convolution, $L$ represents the number of discrete multipaths, $h(n,l)$ is the time-varying complex gain of the ${{l}^{th}}$ path at the ${{n}^{th}}$ sample instant generated from Jakes model\cite{Clarke}\cite{Jakesc}, and $w(n)$ is the additive white Gaussian noise (AWGN). Assuming perfect synchronization at the receiver side, the demodulated signal on the ${{m}^{th}}$ subcarrier in the frequency domain is
\begin{equation}
\begin{aligned}
  Y\left[ m \right]& =W\left[ m \right]+\left( \sum\limits_{l=0}^{L-1}{H_{l}^{0}{{e}^{-j2\pi lk/N}}} \right)X\left[ m \right]\\
 &+\sum\limits_{\begin{smallmatrix}
 k=0 \\
 k\ne m
\end{smallmatrix}}^{N-1}{\sum\limits_{l=0}^{L-1}{X\left[ k \right]H_{l}^{m-k}{{e}^{-j2\pi lk/N}}}}
\end{aligned}
\end{equation}
where $X[k]$ represents the signal transmitted on the ${{k}^{th}}$ subcarrier in the frequency domain, $H_{l}^{m-k}$ represents the FFT of the time-varying multipath channel tap $l$, which also indicates the ICI characteristics between subcarriers given as:
\begin{equation}
\begin{aligned}
H_{l}^{m-k}=\frac{1}{N}\sum\limits_{n=0}^{N-1}{h\left( n,l \right){{e}^{-j2\pi n\left( m-k \right)/N}}}
\end{aligned}
\end{equation}
the second term of (2) indicates the fading coefficient resulting from the multipath except interference of other subcarriers. The third term represents the ICI componet on the ${{m}^{th}}$ subcarrier let
\begin{equation}
\begin{aligned}
\mathbf{X}&={{[X[1],X[2],\cdots ,X[N]]}^T}\\
\mathbf{Y}&={{[Y[1],Y[2],\cdots ,Y[N]]}^T}\\
\mathbf{W}&={{[W[1],W[2],\cdots ,W[N]]}^T}\\
\end{aligned}
\end{equation}
the element of $\mathbf{H}$  in $m^{th}$ row, $k^{th}$ column be $\sum\limits_{l=0}^{L-1}{H_{l}^{m-k}{{e}^{-j2\pi lk/N}}}$. The transmission of an OFDM symbol with $N$ subcarriers can be expressed as:
\begin{equation}
\mathbf{Y}= \mathbf{H}\mathbf{X} + \mathbf{W}
\end{equation}
In our SISO scenario, matrix $\mathbf{H}$ is the frequency domain channel matrix which illustrates the interference and fading to subcarriers in one OFDM symbol.
\begin{figure}
\centering
\setlength{\abovecaptionskip}{0.cm}
\setlength{\belowcaptionskip}{-0.cm}
\includegraphics[width=0.45\textwidth]{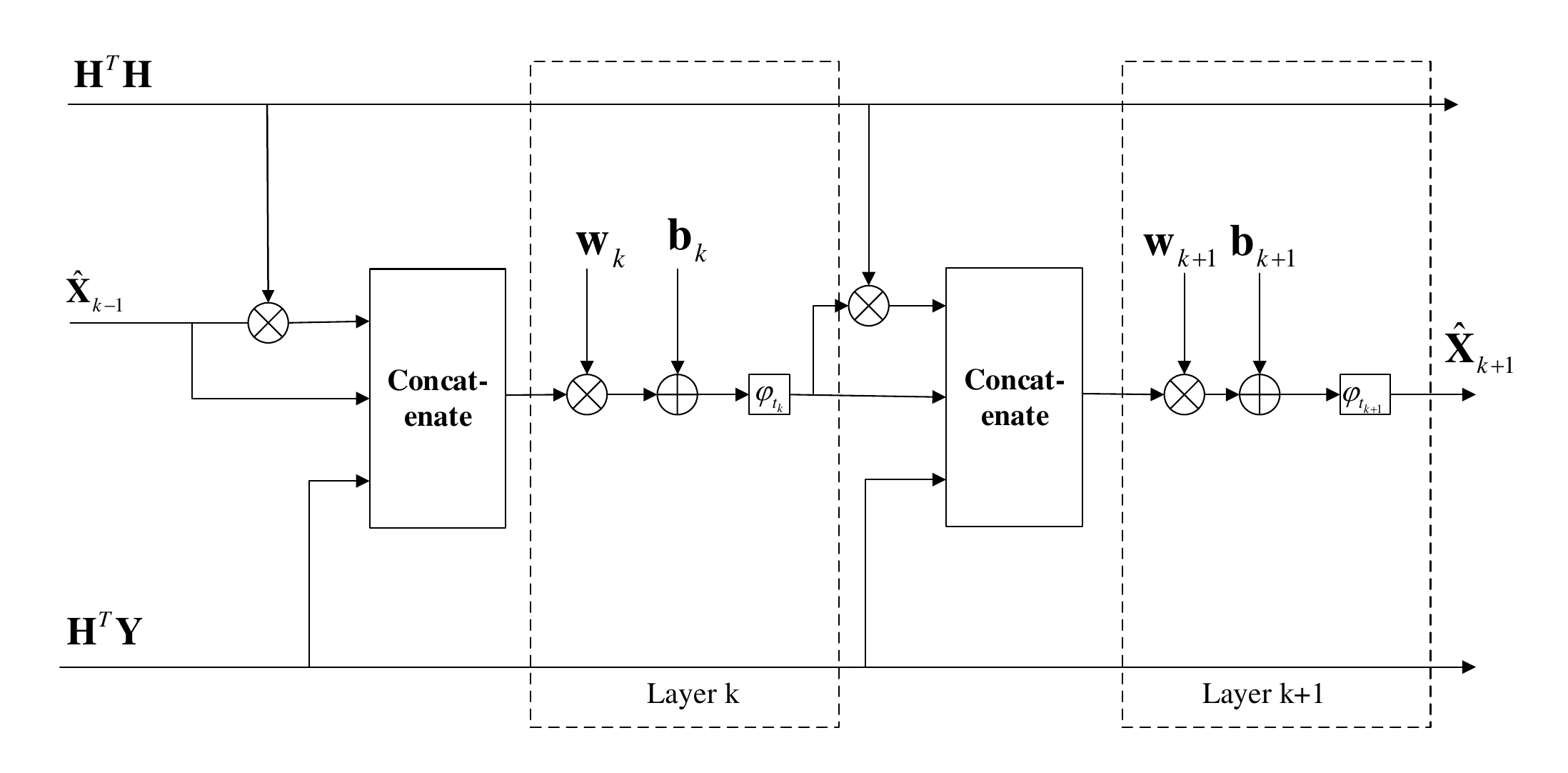}
\caption{The $k^{th}$ layer flowchart of DNT.}
\label{fig:purenet} 
\end{figure}
\subsection{NEURAL NETWORK FOR DETECTION }

In this section, we achieve our deep detection network (DNT) by adding trainable parameters to the traditional detecting algorithm\cite{deepdecode1}. Inspired by article: Learning to detect for MIMO detection\cite{Samuel2018Learning}, we also choose to unfold ML detection algorithm to a trainable network.

The goal in our detection can be expressed in the following equation:
\begin{equation}
\begin{aligned}
{{\mathbf{\hat{X}}}_{\mathbf{\theta} }}(\mathbf{H},\mathbf{Y})=\arg \underset{\mathbf{x}\in {{\{symbolset\}}^{N}}}{\mathop{\min }}\,{{\left\| \mathbf{Y}-\mathbf{HX} \right\|}^{2}}
\end{aligned}
\end{equation}
where the $\mathbf{\theta}$ represents the trainable weights and bias. However, the value of $\mathbf{X}$ is discrete, which is non-differentiable and cannot be optimized. Thus, we enlarge the value set of $\mathbf{X}$ to $\mathbb{C}$ and use hard-decision to achieve the estimation $\mathbf{\hat{X}}$ of sending signal $\mathbf{X}$. Our net's architecture is proposed using deep unfolding\cite{deepunfolding} given as:
\begin{equation}
\begin{aligned}
  & {{{\mathbf{\hat{X}}}}_{0}} = \mathbf{0}\\
  & {{\mathbf{z}}_{k}}={{\mathbf{w}}_{k}}\left[ \begin{aligned}
  & {{\mathbf{H}}^{T}}\mathbf{Y} \\
 & {{{\mathbf{\hat{X}}}}_{k}} \\
 & {{\mathbf{H}}^{T}}\mathbf{H}{{{\mathbf{\hat{X}}}}_{k}} \\
\end{aligned} \right]+{{\mathbf{b}}_{k}} \\
 & {{{\mathbf{\hat{X}}}}_{k+1}}={{\varphi }_{{{t}_{k}}}}({{\mathbf{z}}_{k}}) \\
\end{aligned}
\
\end{equation}
where $\mathbf{\hat{X}_{k}}$ is the estimation of sending signal $\mathbf{X}$ in the $k^{th}$ iteration. $\mathbf{w}_k$, $\mathbf{b}_k$ and $t_k$ are trainable parameters. Intuitively, each iteration is a linear combination. The $k^{th}$ iteration can be seen as the forward propagation from $k^{th}$ layer to ${k+1}^{th}$ layer(see Fig.~\ref{fig:purenet}). After adding trainable parameters. ${{\varphi }_{{{t}_{k}}}}$ is a piecewise linear soft sign activation function cited from\cite{Samuel2018Learning}:
\begin{equation}
{{\varphi }_{t_k}}\left( x \right)=-1+\frac{\rho \left( x+t_k \right)}{\left| t_k \right|}-\frac{\rho \left( x-t_k \right)}{\left| t_k \right|}
\end{equation}
Our net uses a normalized multi-loss function\cite{Samuel2018Learning}, which is :
\begin{equation}
loss\left( \mathbf{X};{{{\mathbf{\hat{X}}}}_{\theta }}\left( \mathbf{H},\mathbf{Y} \right) \right)\text{=}\sum\limits_{k=1}^{L}{\log \left( k \right)\frac{{{\left\| \mathbf{X}-\mathbf{\hat{X}} \right\|}^{2}}}{{{\left\| \mathbf{X}-\mathbf{\tilde{X}} \right\|}^{2}}}}
\end{equation}
where $L$ is the total layer number, $\mathbf{\tilde{X}}$ is the zero-forcing result given as:
\begin{equation}
\mathbf{\tilde{X}}={{({{\mathbf{H}}^{T}}\mathbf{H})}^{-1}}{{\mathbf{H}}^{T}}\mathbf{Y}
\end{equation}
This special designed loss function\cite{Samuel2018Learning} uses zero-forcing detector as a standard to train the network while applying multi loss to prevent network from overfitting. The layer number is same to the number of iteration in origin ML algorithm. Thus, what DNT do is making use of the trainable parameters to find the best detecting algorithm for ML detection in limited iterations. However, as it actually uses all zero vector as the initialization of $\hat{{\mathbf{X}}}_{k}$, training of DNT faces huge difficulty for converging.
\section{CASCADE NET AND SLIDING DETECT ALGORITHM}
\subsection{CASCADE-NET}
\begin{figure}
\centering
\setlength{\abovecaptionskip}{0.cm}
\setlength{\belowcaptionskip}{-0.cm}
\includegraphics[width=0.45\textwidth]{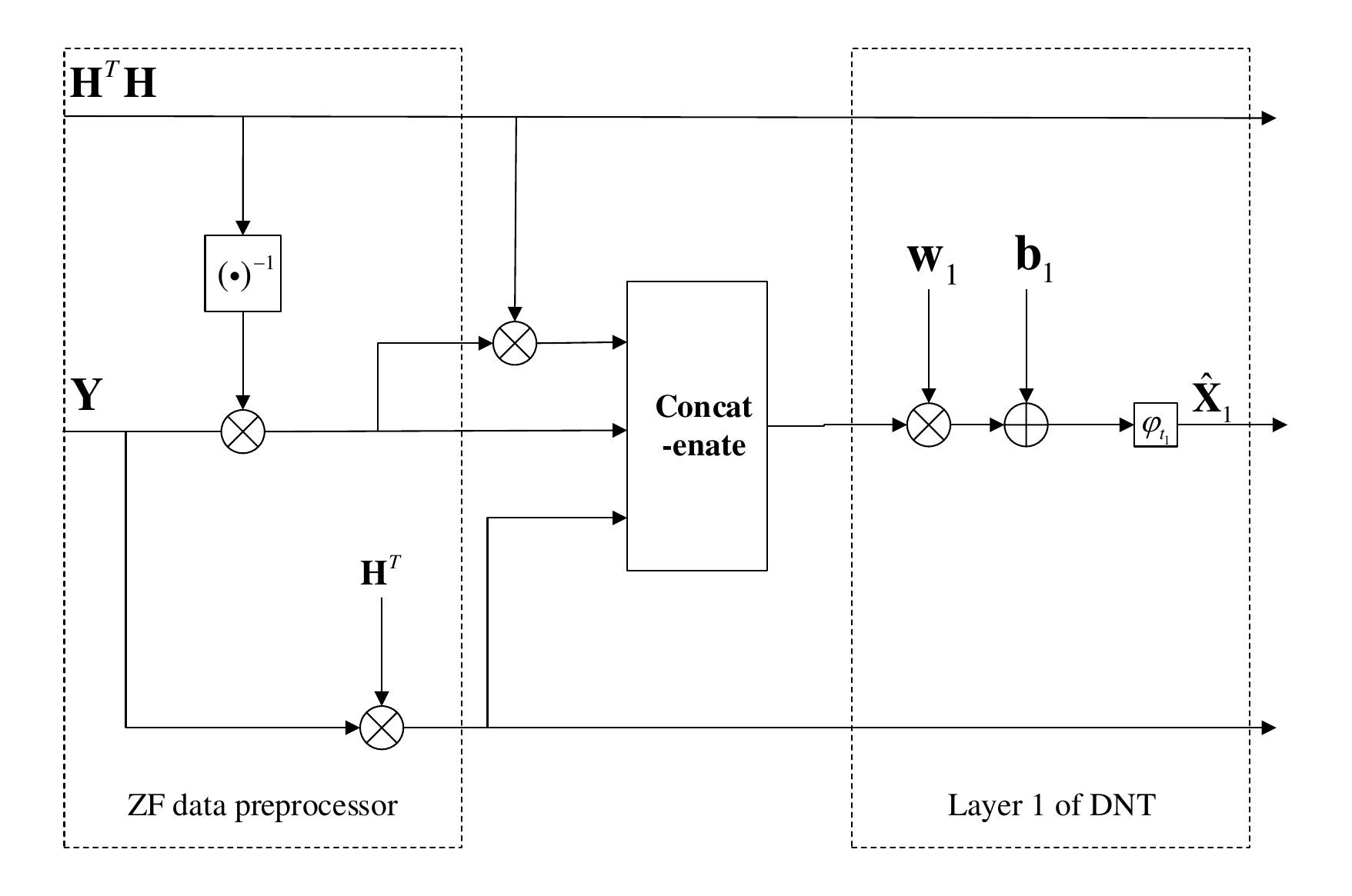}
\caption{The cascade structure.}
\label{fig:cn} 
\end{figure}

As mentioned before, the DNT using zero vector as initialization may suffer from slow convergence. To solve this problem, we proposed a cascade structure. In cascade-net, the DNT is cascaded to a ZF data preprocessor (see Fig.~\ref{fig:cn}). To the DNT, the initializing vector is no longer all zero but roughly processed data providing by ZF detector. In single DNT detector, it has to perform the work of estimating the sending signal from the very beginning, however, in cascade-net its work changes to complete sending signal detection based on the results of ZF detector. In another word, the first part of the cascade-net completes the coarse detection while the second part performs the detailed detection. This idea is inspired by transfer learning. In transfer learning, learning on a target problem is sped up by using the weights obtained from a network trained for a related source task\cite{Transfer}. The parameter fixed network completes parts of work for a new learning target. Similarly, what detectors shared in OFDM symbol detection is restoring sending signal from receiving signal. Thus, cascade-net should have promoted performance on convergence.   
The forward propagation of cascade-net (CN) can be expressed as:
\begin{equation}
\begin{aligned}
  & {{{\mathbf{\hat{X}}}}_{0}} = {{({{\mathbf{H}}^{T}}\mathbf{H})}^{-1}}{{\mathbf{H}}^{T}}\mathbf{Y}\\
  & {{\mathbf{z}}_{k}}={{\mathbf{w}}_{k}}\left[ \begin{aligned}
  & {{\mathbf{H}}^{T}}\mathbf{Y} \\
 & {{{\mathbf{\hat{X}}}}_{k}} \\
 & {{\mathbf{H}}^{T}}\mathbf{H}{{{\mathbf{\hat{X}}}}_{k}} \\
\end{aligned} \right]+{{\mathbf{b}}_{k}} \\
 & {{{\mathbf{\hat{X}}}}_{k+1}}={{\varphi }_{{{t}_{k}}}}({{\mathbf{z}}_{k}}) \\
\end{aligned}
\end{equation}
where ${{\mathbf{\hat{X}}}}_{0}$ is the data preprocessor output and the input of the secondary DNT.

In fact, a cascade-net will not worse than a single zero-forcing detector. If number of subcarriers in an OFDM symbol is $N$ and ${{\mathbf{w}}_{1}}=[{\mathbf{I}_{2N\times 2N}},-{\mathbf{I}_{N\times N}}]$ where $\mathbf{I}$ represents the  identity matrix, substituting ${{\mathbf{\hat{X}}}}_{0}$ into ${{\mathbf{\hat{X}}}_{k}}$ in equation (11) gives:
\begin{equation}
{{\mathbf{\hat{X}}}_{1}}={{\varphi }_{t_1}}[{{\mathbf{w}}_{1}}({{({{\mathbf{H}}^{T}}\mathbf{H})}^{-1}})
{{\mathbf{H}}^{T}}\mathbf{Y})+{{\mathbf{b}}_{1}}]
\end{equation}
Assuming ${\mathbf{w}}_{k}(k>1)$ to be the identity matrix, and ${\mathbf{b}}_{k}$ be the zero matrix, the forward propagation of the cascade-net becomes：
\begin{equation}
{{\mathbf{\hat{X}}}_{k}}={{\varphi }_{{{t}_{k}}}}({{\varphi }_{{{t}_{k-1}}}}\cdots ({{\varphi }_{{{t}_{2}}}}({{\varphi }_{{{t}_{1}}}}({{({{\mathbf{H}}^{T}}\mathbf{H})}^{-1}}{{\mathbf{H}}^{T}}\mathbf{Y})))
\end{equation}
There, we suppose: ${{t}_{k}}={{t}_{k-1}}\cdots ={{t}_{1}}=1$, ${{\varphi }_{{{t}_{k}}}}(x)$ satisfies ${{\varphi }_{{{t}_{k}}}}(x)=x$ in the activation area of the function. As the constellation points have already been normalized before transmitting, their imaginary parts and real parts are less than $1$. Therefore, nonlinear opponents will not affect the final hard-decision of the constellation point.
At this time, the output of the cascade-net equals to a single zero-forcing detector. In another word, zero-forcing detector is one of the solutions to the training of this net. Now, we can conclude that by cascading a zero-forcing detector as a data preprocessor, the training mission changes to find a better detector with the basis of zero-forcing detection.

In the DNT, a normalized loss function is used for evaluation. However, in cascade-net this operation is redundant, as the first level is zero-forcing detector, the cascade-net natively consider the zero-forcing results as a standard, the optimization of cascade-net is continuing learning from zero-forcing detection to ML detection. Thus, in cascade-net, we directly use Euclidean distance as the loss function, which is:
\begin{equation}
loss\left( \mathbf{X};{{{\mathbf{\hat{X}}}}_{\theta }}\left( \mathbf{H},\mathbf{Y} \right) \right)\text{=}\sum\limits_{k=1}^{L}{\log \left( k \right){{\left\| \mathbf{X}-\mathbf{\hat{X}} \right\|}^{2}}}
\end{equation}
\subsection{SLIDING STRUCTURE}
\begin{figure}
\centering
\setlength{\abovecaptionskip}{0.cm}
\setlength{\belowcaptionskip}{-0.cm}
\includegraphics[width=0.45\textwidth]{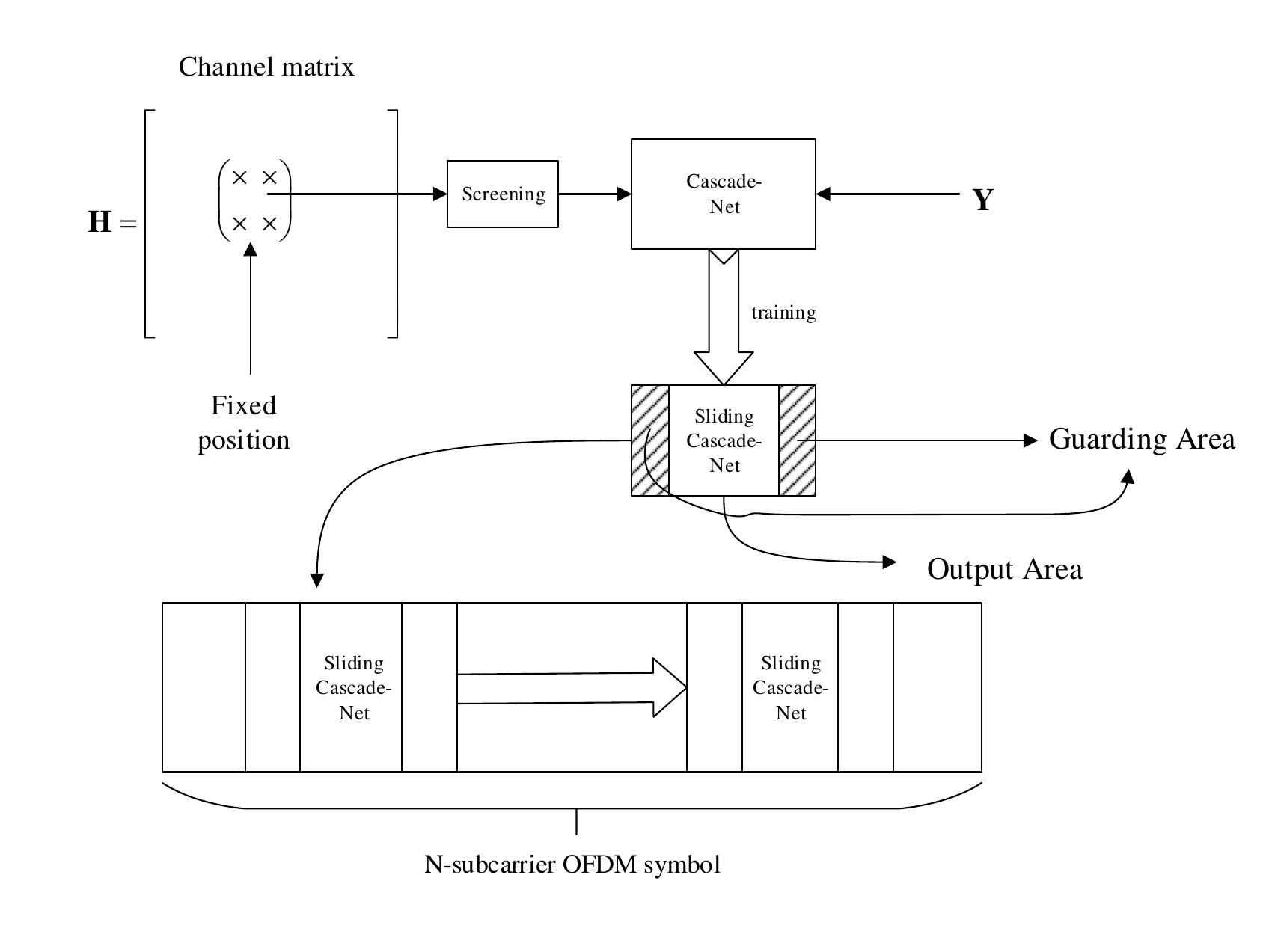}
\caption{The structure and application of SCN.}
\label{fig:SN} 
\end{figure}

In modern communication system, number of subcarriers in one OFDM symbol can be very large. Therefore, using one cascade-net to learn the detection of such a OFDM symbol requires dramatic calculation resource which is hard to be realized.
However, ICI between subcarriers has strong correlation among them which indicates the possibility of peforming the whole detection step by step. In fact, ICI to one subcarrier is mainly caused by limited number of adjacent subcarriers, therefore we propose a sliding detecting structure\cite{fenkuai} for detection of those OFDM symbols with large number of subcarriers. Different from the classical ICI cancellation method\cite{zhao}, sliding window (SW) consists of two parts: guarding area (GA) and output area (OA). GA is used to help OA complete the ICI cancellation, which means the detection result in this part will not be output. Then, SW keeps sliding until finishing the detection of the whole symbol. The structure of SW and the detecting process are shown in Fig.~\ref{fig:SN}. Design of these two parts is based on the signal to interference power ratio and feasibility of calculation.

In previous section, we introduce cascade-net for OFDM detection. Sliding cascaded-net is applying sliding window to cascade structure, which means each SCN will consist of two parts: zero-forcing data preprocessor and deep detection network. The training to SCN is finding proper parameters to optimize the detection performance with submatrix of $\mathbf{H}$ and $\mathbf{Y}$ as feed-in data. We treat each SCN as a CN for OFDM symbol detection with $l_T$ subcarriers where $l_T$ correspond to the length of SCN.
After proper training, SCN detects $l_T$ subcarriers in each slide and keeps sliding until it completes detection of the receiving OFDM symbol shown in Fig.~\ref{fig:SN}.
Suppose $\mathbf{\hat{X}}_{\theta }^{scn}(\mathbf{H},\mathbf{Y})$ to be the converged SCN detector with proper training. After $n_{slip}$ times sliding, the symbols to be detected ${{\mathbf{Y}}_{{{n}_{slip}}}} $ , and the SCN detector output $\mathbf{X}_{{{n}_{slip}}}^{out}$ can be expressed as:
\begin{equation}
\begin{aligned}
  & {{\mathbf{Y}}_{{{n}_{slip}}}}=\left[ {{Y}^{({{n}_{slip}}-1)l_{T}}}{{Y}^{({{n}_{slip}}-1)l_{T}+1}}\ldots {{Y}^{{{n}_{slip}}l_{T}}} \right] \\
 & {{\mathbf{X}}_{{{n}_{slip}}}}= \mathbf{\hat{X}}_{\theta }^{scn}(\mathbf{H}_{n_{slip}},\mathbf{Y}_{n_{slip}}) \\
 &\mathbf{X}_{{{n}_{slip}}}^{out}=[X_{{{n}_{slip}}}^{l_{G}+1}X_{{{n}_{slip}}}^{l_{G}+2}\ldots X_{{{n}_{slip}}}^{l_{T}-l_{G}}] \\
\end{aligned}
\end{equation}
where  $l_G$ correspond to the length of GA in SCN, $\mathbf{H}_{n_{slip}}$ represents the sub-matrix used in ${n_{slip}}$-times detection. GA includes those subcarriers which corresponds to main interference to subcarriers in OA. Therefore, GA is used to assist OA completing ICI cancellation.

Next, we give out the specific details about the GA design. In fact, interference of $k^{th}$  subcarrier to the $m^{th}$ is a decreasing function.
It indicates that interference to a specific subcarrier concentrates on limited number of adjacent subcarriers. Based on this idea, the empirical formula we proposed is:
\begin{equation}
\begin{aligned}
  & {{x}_{l}}=\min \{\underset{{{x}_{l}}>{{f}_{N}}}{\mathop{\arg }}\,\{\frac{1}{{{N}^{2}}}\cdot {{\left\| \frac{\sin \pi {{x}_{l}}}{\sin \pi {{x}_{l}}/N} \right\|}^{2}}=\alpha\beta \}\} \\
 & {{l}_{G}}=\left\lceil {{x}_{l}}-{f_{N_d}} \right\rceil  \\
\end{aligned}
\end{equation}
where $\beta$ indicates the lowest interference power considered in one SW while $\alpha$ is used to compensate for unequal amplitude modulation. The length of output area mainly depends on the capability of the calculation resource.

To SCN  the signal in GA is used to assist the detection of signal in OA. In another word, we actually do not care about the output of GA. Thus, the loss function should also only focus on the performance of OA. Based on this idea, the loss function used to train SCN is now promoted to this form:
\begin{equation}
loss({{\mathbf{X}}_{out}};\mathbf{\hat{X}}_{\theta }^{scn}(\mathbf{H},\mathbf{Y}))=\sum\limits_{k=1}^{L}{\log (k)}{{\left\| {{\mathbf{X}}_{out}}-{{{\mathbf{\hat{X}}}}_{out}} \right\|}^{2}}
\end{equation}
where ${\mathbf{X}}_{out}$ and ${{\mathbf{\hat{X}}}}_{out}$ represents the signal in output area.

\begin{table}
    \caption{The parameters used for detection.}
    \label{tab:parameters0}
    \centering
    \begin{tabular}{lllll}
        \toprule[1pt]
        &Scenario: &$N=32$  &$f_{N_d} = 0.16 $ or $0.18$ \\
        \midrule
        Label&Modulation&Learning Rate&Layers\\
        Value& QPSK & 0.005&20 \\
        Label&BatchSize\\
        Value&500\\
        \bottomrule[1pt]
    \end{tabular}
\end{table}
\section{LEARNING TO DETECT AND NUMERICAL RESULTS}
In this section, we compare the detection performance of CN to DNT and classical ZF detector. The deep detectors are trained off-line and applied online. Our simulation is based on the assumption that receiver can get accurate channel information. To prevent our network from miss adjustment, we dismiss the channel matrix with condition number larger than $10000$ in training phase. The signal to noise rate (SNR) of the data in training set depends on the range of SNR while detecting. If the detecting range is $15dB$ to $35dB$, the SNR used for training should be $\frac{15+35}{2} = 25dB$. The trainable parameter $\mathbf{w}_k$ is initialized using truncated norm function with mean $0$ and  variance $1$.
$\mathbf{b}_k$ is initialized with $0.01$, and $t_k$ is initialized with $1$. We use tensorflow to construct our network, and apply Adam algorithm as optimizer. TABLE~\ref{tab:parameters0} shows the parameters for the detection of OFDM symbols with $32$ subcarriers transmitted in $4$ paths fading channel\cite{lte} with normalized Doppler shift ${f_{N_d}}=0.16$ or $0.18$ constructed by Jakes model.
\begin{figure}
\centering
\setlength{\abovecaptionskip}{0.cm}
\setlength{\belowcaptionskip}{-0.cm}
\includegraphics[width=0.41\textwidth]{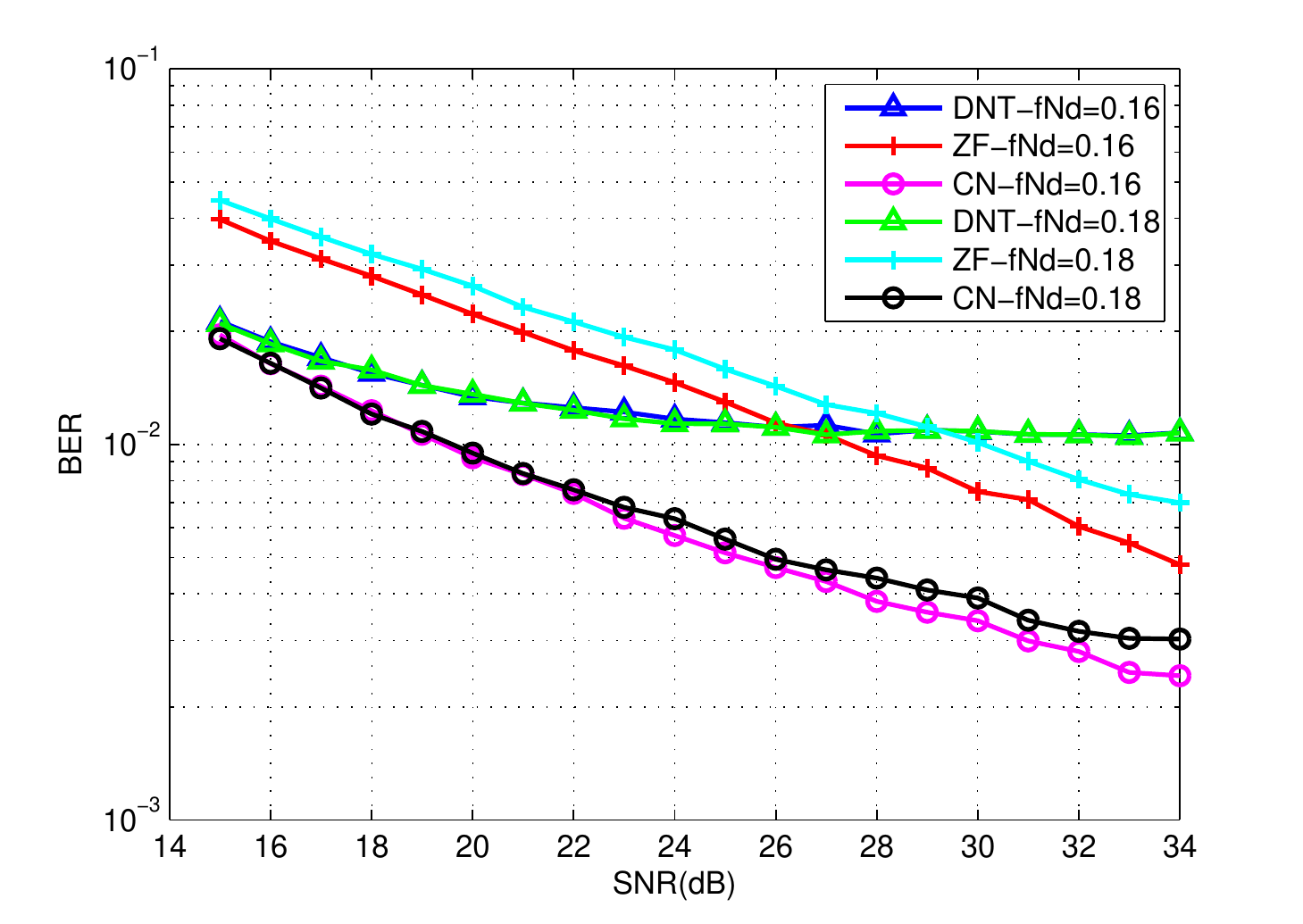}
\caption{BER versus SNR of 32-subcarrier OFDM symbol detection on QPSK with $f_{N_d}=0.16$ or $0.18$}
\label{fig:ofdm32} 
\end{figure}

The result shown in Fig.~\ref{fig:ofdm32}. illustrates that when signal-to-noise ratio is around $20dB$, the performance of DNT is much better than classical zero-forcing detector, however when SNR goes up the constrained iteration leads to the decline of DNT's performance. To CN, due to the first level of data preprocessor, it always achieves a better performance than zero-forcing detector.

In SCN, the calculation complexity of matrix inverse would be $\Theta \left( {{S}^{3}} \right)$, while the complexity of sliding detector is linear. Thus, the output part should be as short as possible.
To reach a compromise between complexity and detection delay, we suggest the length of the output part $l_O$ to be $8$ or $16$. The total length $l_T$ of SCN equals $l_O+2l_G$. The total sliding times per detection is $\frac{N}{l_O}$. One more thing should be taken into consideration in SCN training is the selecting of sub-matrices. We get the submatrices $\mathbf H_T$ from the fixed location $p$ to $p+l_T$ in channel matrices $\mathbf H$ (shown in Fig.~\ref{fig:SN}).
\begin{table}
    \caption{The parameters used for detection.}
    \label{tab:parameters1}
    \centering
    \begin{tabular}{lllll}
        \toprule[1pt]
        &Scenario: &$N=256$  &$f_{N_d} = 0.16$ or $0.18$\\
        \midrule
        Label&Modulation&Output Area&Guarding Area \\
        Value& QPSK & 16&8 \\
        Label&Learning Rate &Layers&BatchSize\\
        Value&0.005 &20&1500\\
        \bottomrule[1pt]
    \end{tabular}
\end{table}
This method is practical as the statistical characteristic of the channel is ergodicity. TABLE~\ref{tab:parameters1} shows the hyper-parameters for the detection of OFDM symbols with subcarrier $N=256$ transmitted in the same fading channel as above. The GA length is designed using the method mentioned equation (16).
\begin{figure}
\centering
\setlength{\abovecaptionskip}{0.cm}
\setlength{\belowcaptionskip}{-0.cm}
\includegraphics[width=0.41\textwidth]{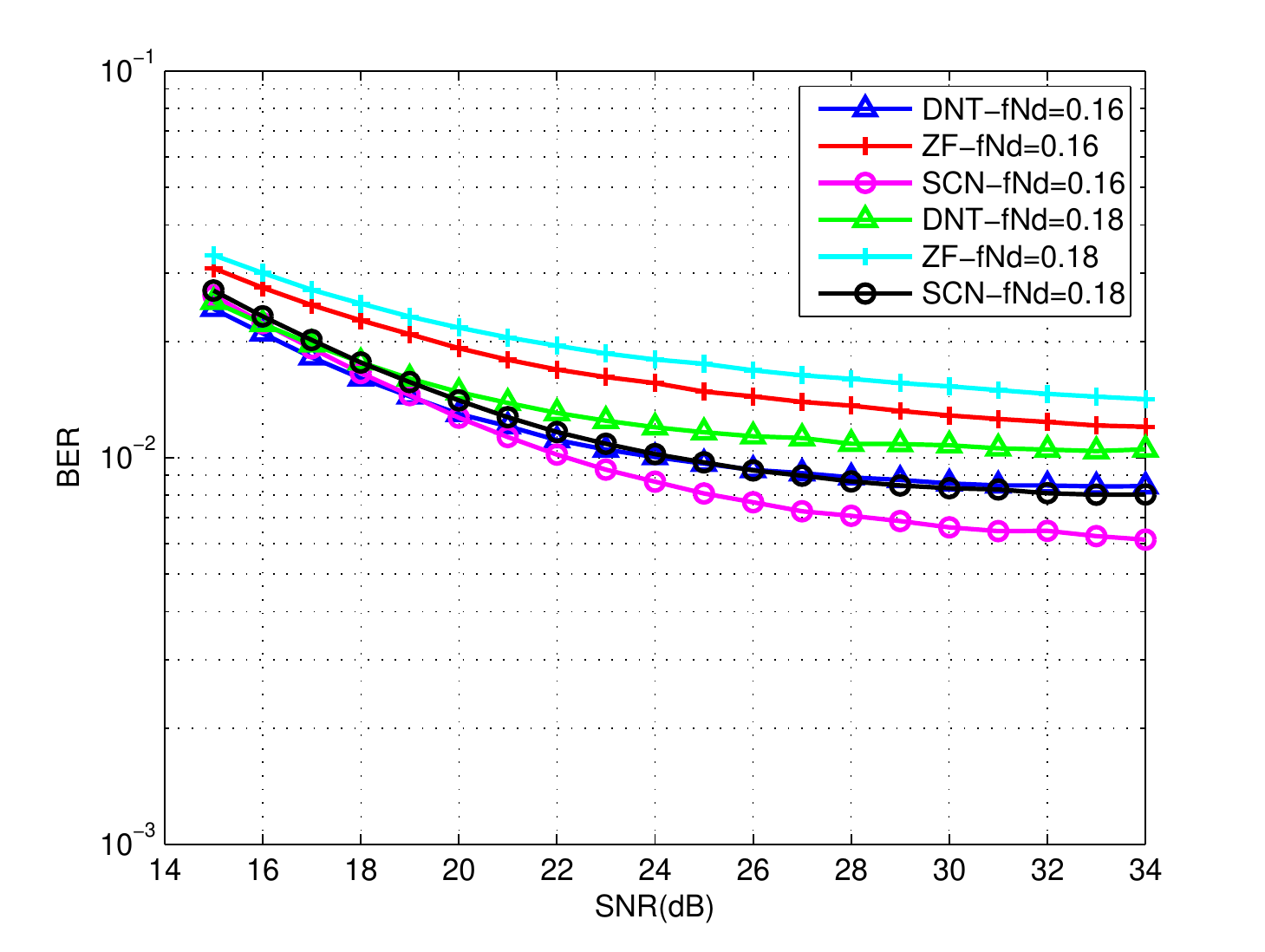}
\caption{BER versus SNR of 256-subcarrier OFDM symbol detection on QPSK with $f_{N_d}=0.16$ or $0.18$}
\label{fig:ofdm256} 
\end{figure}
The results in Fig.~\ref{fig:ofdm256}. shows that both classical sliding detector and deep sliding detector have the error floor. However, after using deep sliding detector, the detector performance is obviously promoted. SCN makes further promotion because the error floor of it is the lowest.

\section{Conclusion}
In this paper, we propose a cascade network for detection of OFDM symbol transmitted in channel with large Doppler shift and provide a sliding structure for the detection of multi-subcarrier OFDM symbol. Simulations based on QPSK indicate that our network performs better than classical zero-forcing detector. Moreover, it has a better performance than single deep detection network in high signal-to-noise ratio. Though perfect channel information is assumed in our simulation, cascade-net is robust against inaccurate channel estimation. In fact, sliding structure could also be used in the first level (data preprocessor) of SCN for further reduction of computational complexity of data preprocessor.
\section{Acknowledgment}

%
%



\bibliographystyle{IEEEtran}
\bibliography{Deep_detect}
%
%
%

\end{document}